# Towards a 2D Printer: A Deterministic Cross Contamination-free Transfer Method for Atomically Layered Materials

Rohit A. Hemnani[1], Caitlin Carfano[1], Jason P. Tischler[1], Mohammad H. Tahersima[1], Rishi Maiti[1], Ludwig Bartels[2], Ritesh Agarwal[3], Volker J. Sorger[1*]

[1]Department of Electrical and Computer Engineering, George Washington University, Washington, DC 20052, USA
[2]Chemistry and Materials Science and Engineering, University of California, Riverside, California 92521, USA
[3]Department of Materials Science and Engineering, University of Pennsylvania, Philadelphia, PA 19104, USA
*Corresponding Author E-mail: *sorger@gwu.edu*

**Abstract**
**Precision and chip contamination-free placement of two-dimensional (2D) materials is expected to accelerate both the study of fundamental properties and novel device functionality. Current transfer methods of 2D materials onto an arbitrary substrate deploy wet chemistry and viscoelastic stamping. However, these methods produce a) significant cross contamination of the substrate due to the lack of spatial selectivity b) may not be compatible with chemically sensitive device structures, and c) are challenged with respect to spatial alignment. Here, we demonstrate a novel method of transferring 2D materials resembling the functionality known from printing; utilizing a combination of a sharp micro-stamper and viscoelastic polymer, we show precise placement of individual 2D materials resulting in vanishing cross contamination to the substrate. Our 2D printer-method results show an aerial cross contamination improvement of two to three orders of magnitude relative to state-of-the-art dry and direct transfer methods. Moreover, we find that the 2D material quality is preserved in this transfer method. Testing this 2D material printer on taped-out integrated Silicon photonic chips, we find that the micro-stamper stamping transfer does not physically harm the underneath Silicon nanophotonic structures such as waveguides or micro-ring resonators receiving the 2D material. Such accurate and substrate-benign transfer method for 2D materials could be industrialized for rapid device prototyping due to its high time-reduction, accuracy, and contamination-free process.**

**Introduction**
Two-dimensional (2D) materials hold promise for atomic-scale and highly-functional electronic and photonic devices by taking advantage of their rich physical properties which are fundamentally different from their bulk counterparts [1-7]. The diverse bandgap range of atomically-layered materials include insulators hexagonal Boron Nitride (hBN), semiconductors ($MoS_2$, $MoSe_2$, $WS_2$ etc.), and semimetals (graphene) providing the opportunity to realize heterostructures without the conventional lattice mismatch known from, for instance, integrating III-V materials with group IV substrates [8-17]. Graphene, the first member of the 2D family, shows semi metallic electronic properties and has proven to be a key active material for a wide range of applications including in the optoelectronic industry [2,3]. On the other hand, semiconductor transition metal dichalcogenides (TMDs) and black phosphorus demonstrate thickness dependent bandgap tunability covering the visible and near infrared spectrum and direct bandgap with high emission yields [18-22]. Additionally, hBN is an ideal insulator for 2D material heterostructure platform given its bandgap of ~5.9 eV and low break-down voltage [11-13]. Strong in-plane covalent bonds provide stability of 2D crystals, whereas the vertical layers of 2D material are held together by weak Van der Waals forces and can be easily separated through the mechanical exfoliation method [22]. In contrast, chemical vapor deposition (CVD) offers chip-scale 2D material growth relevant for industry [23]. However, in addition to quality imperfections, both the thermal budget and substrate incompatibility currently limit CVD growth methods, despite significant progress having been made [24]. Hence, the scalability of obtaining these 2D materials and the ability to reliably transfer a 2D material flake onto an arbitrary substrate with minimum cross-contamination remains challenging. The capability to transfer, for instance, a single flake to a target area is important because neighboring features such as other devices, circuits, or photonics waveguide structures, could be harmed if the 2D material flake areas are randomly introduced onto the target substrate. Previous transfer approaches, such as the wedging method use wet chemistry to transfer 2D materials [25]. Whereas, the polymer (evalcite) transfer method involves the substrate heating and wet chemistry to obtain a successful transfer [26]. As described, currently transfer methods do not encompass the precautions to protect neighboring devices; this could become critical for heterogeneous integration such as carried out in Silicon photonics [27,28]. With each method, there is a drawback that could harm a specific electronic or photonic device, whether it be chemicals, heating, excess residue, etc. Here, we demonstrate a novel method to transfer a precisely placed 2D material onto an arbitrary substrate with vanishing (<99%) cross-contamination. To show the added advantages of this method, we next compare three currently used approaches, that are simple to replicate and do not involve heating to our new one; these comparisons are the Direct [29], Dry [30], and our proposed Litho-Assisted, and our micro-stamper-based method, termed here '2D-Printer' method. Using the latter, we demonstrate how time-efficiently 2D materials can be placed on various electronic devices using the manual 2D material printer within a virtually zero cross-contamination range. We describe methodology on future improvements to industrialize the printer with rapid optical identification technique [31-33] and machine learning to increase scalability for various practical applications.

**Methods and Results**
Before discussing the performance of our 2D printer, we briefly state the conceptual methodology of the Direct, and Dry transfer methods to provide a holistic picture of the steps involved in each. The Direct transfer approach involves placing a bulk crystal between two pieces of adhesive tape (Scotch or Nitto tape) and then peeling one tape off the other (Fig. 1a) [22]. This action breaks weak Van der Waals interactions keeping the layers of crystal together and essentially thins it down. This procedure is repeated numerous times until ideally a flake with desired thickness is obtained. Then, we bring scotch tape with the exfoliated flakes in direct contact with the substrate (Fig. 1a). This simple and quick direct transfer approach always offers optimal quality layered materials. However, one major limitation is to control the thickness and lateral dimensions of the flakes. The obtained flakes differ considerably in size and thickness, where the sizes range from nanometers to several tens of micrometers for single to few layers 2D materials.

In the Dry Transfer approach [29], one uses an intermediate polymer PDMS gel instead of directly placing the same scotch tape with exfoliated flakes onto the substrate. Afterwards, we place the tape in direct contact with the PDMS gel and peel off the tape rapidly. The flakes adhered to the gel and the gel was placed onto a glass slide. On the other hand, the glass slide with the gel was attached to the three-axis micromanipulator stage and placed near a three-axis micromanipulator stage holding the substrate as shown in Figure 1(b). Since the PDMS gel is transparent, the axis of micromanipulators of the desired flake can be directly placed over the target area on the substrate of our choice easily using a microscope. This transfer method relies on the viscoelastic properties of the PDMS stamp where the gel is peeled off slowly from the substrate. This allows the flakes to adhere to the substrate rather than the gel because the strength of adhesion is dependent on the speed at which the gel is pulled [29, 34]. This transfer method is advantageous because a flake can be precisely placed; however, it is not ideal considering that not only the desired flake will be transferred since the flat core slide is pressed onto the substrate in its entirety. Thus, a non-finite probability exists that some other flakes are transferred (Fig. 2b). In dry transfer and the direct transfer, we basically create the imprint of the source. Theoretically, dry transfer can also lead to cross contamination free transfer if we have a single flake source however in reality scotch tape exfoliation results in flake clustering, which makes it difficult to find an isolated flake. In order to decrease this random cross-contamination further, the third transfer method, i.e. Litho-Assisted approach has been tested.

Here, the transfer method requires the substrates to be coated with polymethylmethacrylate (PMMA) and an opening developed on a desired location using electron beam lithography (EBL) as shown in Figure 1(c). Theoretically, after the flakes have been transferred, flakes will remain inside the target area within the box as well as randomly distributed on top of the PMMA. Upon washing off the PMMA (i.e. Acetone rinse) the access 2D flakes residing on the polymer are taking off as well, and thus this step resembles a lift-off process. To improve further, the 2D Printer Transfer (Fig. 1d) is implemented, which works as follows; a PDMS gel is stretched over the substrate and a flake is lined up with the target area. Here, a micro-stamper is used to bend the PDMS to push the flake underneath onto the target area. This method gives the same alignment precision as the Dry transfer, which determined by the overlay accuracy of the micro-stamper, 2D material and substrate. Theoretically, with this method, only one single flake will be transferred exactly to the target area. This novel idea of using an intermediate polymer (PDMS) along with a micro-stamper pertain to a significantly reduced cross-contamination rate.

Next, we quantitatively compare the cross contamination ($CC$) of the four different transfer approaches (Fig. 2). In order to quantify the amount of cross contamination for each of these four transfer methods (Direct, Dry, Litho-assisted, 2D-Printer, Fig. 1), we optically measure the sum of all undesired transferred flakes ($A_{access}$) in a certain predefined chip area ($A_{box}$). Here, $A_{box}$ is a multiple ($f$ = 100x, 400x, and 900x) of the flake area ($A_{flake}$) in order to account for varying flake sizes (i.e. $A_{box} = f \times A_{flake}$). Then the cross contamination is obtained via

$$CC = \left( \frac{\sum A_{access}}{A_{box}} \right)_f \quad (1)$$

Repeating the transfer experiment for each method 10 times, we can fit the resulting distributions with a Gaussian function, since each experiment is an independent sample. The resulting mean values ($\bar{\mu}$) and standard deviations ($\sigma$) from the fitting show that the spatial cross contamination spread is most significant when considering small areas around the target transferred flake (i.e. $A_{box}$ = 100x) (Fig. 2e). This can be understood by the spatial distribution of the 2D source use for the transfer method; the repeated mechanical stamping (often manual pressing) on a similar location using scotch-tape during exfoliation leads to clustering of 2D materials. Thus, both the Direct and the Dry transfer methods show an anticipated *CC* improvement with larger $A_{box}$, highlighting the 'edge' of the cluster area.

Our results show the highest amount of cross contamination for the Direct transfer as there is no controlled placement used (e.g. $\bar{\mu}$ = 50%, $f$ = 100x, Fig. 2a,e). In addition, we observe a wide distribution in the cross-contamination histogram signifying the randomness of this approach (standard deviation = $\sigma$ = 35%). This method also leaves much residue from the tape on the substrate, which is undesired. However, the Dry Transfer method shows a narrower distribution (e.g. $\bar{\mu}$ = 28%, $\sigma$ = 8%, $f$ = 100x, Fig. 2b,e), due to the ability to select a region of the PDMS/cover slide. In addition, since the entire centimeter-large cover slide touches the substrate, a non-finite probability exist that other structures or sensitive device regions are either contaminated or physically damaged during the transfer. In contrast, we observe a significant improvement when using the PMMA coated substrate with electron beam lithography (EBL) pre-transfer fabricated openings (i.e. Litho-Assisted method, Fig. 2c). This ensures that any unwanted flakes on the gel are not transferred to the substrate directly as blocked by the PMMA coating. Although varying amounts of cross-contamination persist, this process is an improvement over the dry transfer by a factor of 15 defined as the (Fig. 2c,e). For areas larger than 100x of the flake area, the cross contamination improves by a factor 22 (Fig. 2e). This method is an improvement over the Dry transfer method due to the PMMA greatly reducing the cross-contamination, the 2D Printer reduces cross-contamination most significantly (Fig. 2d,e). This method shows a virtually cross contamination-free transfer by two order of magnitude from the Dry transfer (Fig. 2d). This is surprising as we expect the spatial accuracy for the Litho-assisted and micro-stamper-based 2D printer method to be similar, but can be attributed to the fact that when transferring a flake within a fabricated opening, the flake has a chance of partially draping on top of the PMMA. Once the PMMA is lifted off, some of the flake will be lifted off with it, while the remainder has a probability to become fragmented and left behind on the substrate.

Nonetheless, this method is not holistically viable since PMMA could damage a chemically-sensitive device. Additionally, trying to stamp on multiple openings on the same device poses a problem because there is a chance that flakes will transfer in adjacent openings rather than only the desired opening. To circumvent these issues, the 2D printer method is developed as introduced in this work. Upfront, without the need for a PMMA substrate coating, it encompasses the full range of devices as well as a reduction in cross contamination. For instance, the mean ratio of the 2D printer vs. Direct and Dry methods for the smallest test region are 310 and 170 respectively. The randomness of the latter two methods is somewhat mitigated when the test area is increased (e.g. 950 and 700).

The 2D printer consists of a two-axis stage micromanipulator holding the substrate (Fig. 3a). The second stage is a three-axis micromanipulator which holds clips responsible for holding the thin PDMS gel. Another three-axis micromanipulator is used to control the micro-stamper. Multiple micromanipulators are needed because the micro-stamper placement must be very precise- a slight shift of a few micrometers could drastically change the flake alignment. While improved equipment such as high magnification microscope objective lenses with (ideally) long working distance and nanometer or micrometer precision mechanical components yield better overlap accuracy, the setup used in this work has a lateral transfer resolution of about ±5 μm, mainly limited by mechanical precision and optical resolution. The latter sets a bound given by the long required working distance of the objective lens. In addition, it is necessary to consider the size of the flake being transferred. If a flake is smaller than the micro-stamper, the flake will transfer ?100% of the time because the entire surface area of the flake touches the substrate. On the other hand, if a flake is larger than the micro-stamper, the flake will not transfer reliably. In this case, either a part of large flake will transfer or there will be no transfer at all due to the multilayer nature. only Since the micro-stamper and the PDMS are key factors in reducing cross contamination, testing is further done to ensure accuracy when the experiment is reproduced, we find that when the micro-stamper is in contact with the substrate, the PDMS acts as a shock absorber, preventing any possible damage to the device. Raman measurements have been performed to investigate the quality of the 2D flakes before and after the transfer process as depicted in Figure 3(b). Looking at the spectra for both before and after transfer, the peaks rise (mention the peaks here) at the same point ensuring that the material's identity has not been affected in this transfer process. Industrial application for Van der Waal heterostructures will require a scalable approach to stack 2D materials on top of each other and an arbitrary substrate with any morphology. This 2D Printer method is reliable both in theory and practice and can be further improved by autotomizing rather than manual operation. The apparatus in Figure 3(a) could be automated with a motor using a microcontroller or Raspberry Pi to rotate both stages according to user input, effectively reducing human error of misalignment. One of the major characteristics of these exfoliated flakes is a correlation between the color of the flakes taken with an optical white light microscope images versus the heights of the flakes [28-30]. AFM height measurement data will be taken and fed into a machine-learning algorithm to determine flake heights that were not explicitly determined. The color contrast will be calculated and a data spreadsheet will be created corresponding the color and height variables. A user interface will ask for desired 2D material, desired flake height, desired flake size, and target area for input. Such rapid optical identification algorithm will make it possible to quickly locate the desired characteristics and align the two micromanipulator stages and the micro-stamper will be brought down automatically. This will create an industrial style quick, efficient, and low cross-contamination 2D material printer for use of placing 2D materials on optoelectronic devices, photonic integrated circuits, etc.

The advantages of the 2D material printer are highlighted in Figure 4 showing various practical applications. For instance, we demonstrate the capability to successfully transfer TMDs on an array (5x2) of boxes on a silicon substrate shown in Figure 4(a). There is little to no cross contamination on the device, providing the ability to have multiple stampings on the same substrate to create working devices. This is contrasted to using the Dry transfer method for this task as depicted in Figure 4(b). The cross contamination of the device is so significant that the device is useless. Figure 4(c) presents the integration of h-BN on Si waveguide by our 2D printer method to demonstrate the damage-less transfer. We have transferred few layer h-BN on a linear Si waveguide multiple times and measured the propagation loss each time. The measured waveguide output power remains unchanged, just limited by the noise our experimental setup (inset Figure 4c). This confirms that the 2D Printer is a viable approach for realizing heterogeneous integrated photonic device applications. We have demonstrated the fabrication of Van der Waals heterojunction as depicted in figure 4(d). Due to the precision placement and cross-contamination-free essence of this method, it can be applied for the creation of multiple, quick and accurate heterostructures on the same device. This advantage is further exemplified by another example application of two different processes by transferring a TMD flake onto a Si micro-ring resonator (Fig. 4e, f). As expected, we observe, a high-degree of cross contamination for the Dry method with flakes scattered across the silicon photonic waveguides and covering neighboring devices, and thus renders this method unusable for real chips. Repeating the experiment with the 2D Printer method shows that only the single targeted flake is placed onto the micro-ring resonator.

**Conclusion**

In summary, we have developed and demonstrated a novel transfer method for 2D materials utilizing a micro-stamping technique to significantly reduce the lateral cross contamination area on the substrate receiving the 2D material, and improving spatial accuracy. Here, we experimentally demonstrated that using a conventional micro-stamper can significantly improve the transfer of multilayer and few layer 2D materials—reliably reducing cross contamination often caused by other transfer methods. Compared to the widely used Dry transfer method, our 2D printer method shows a virtually cross contamination-free (>99% clean) transfer methodology. This capability will significantly increase the range of application of 2D materials. We also demonstrate the diversity of applications that this method can perform (i.e. electronics, photonics, plasmonics, on-chip circuits, etc.) as well as showing that it does not damage these devices. Additionally, this method can be easily automated by combining rapid optical identification algorithms along with simple motors. The manual printer can be further improved by these means of automation, reducing the human error as well as improving transfer speed. This fast and efficient method will provide a means for expedited research regarding 2D materials on optoelectronic devices, heterostructure fabrication, and more.

**Acknowledgements**
VS is supported by ARO (W911NF-16-2-0194) and by AFOSR (FA9550-17-1-0377).

# Manuscript Figures

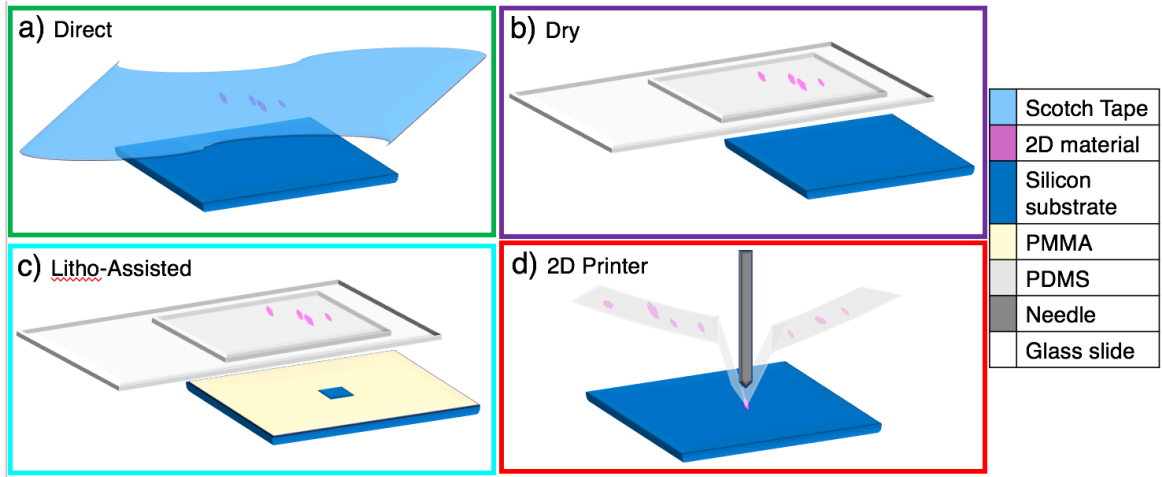

**Figure 1.** Schematic illustration of different dry transfer methods of 2D materials investigated and compared in this work. (a) Schematic of the direct transfer method (Direct). A piece of scotch tape with exfoliated flakes on it is directly placed onto the substrate. (b) Schematic of deterministic dry transfer method. The exfoliated flakes are transferred to a PDMS gel from the scotch tape and the gel is then aligned with the substrate. Since the PDMS is transparent, it is possible to see both the flake on the gel and the target area on the substrate simultaneously. Both the gel and the substrate are on micromanipulator stages, once the flake is aligned with the target area, we lower the gel to the substrate. We then lift off the gel slowly, leaving the flakes on the substrate. (c) Schematic of lithography- assisted deterministic dry transfer method (Litho-Assisted). The same steps of the Dry transfer method are followed; however, the substrate is coated in PMMA and an opening created by using Electron Beam Lithography (EBL). When the PDMS gel is push down on the PMMA coated substrate, all the flakes in contact with the PMMA/substrate will transfer but only the flakes that are within the EBL opening will stay. The rest will get washed off with acetone along with the PMMA similar to the conventional lift-off process. (d) Schematic of the 2D printer method (2D Printer). A micro-stamper is used to pinpoint the exact location the flake is to be transferred onto the substrate. A PMMA coated substrate with EBL opening substrate is used and the PDMS gel is placed directly over the substrate at a distance and only the part with the flake comes into contact with the substrate by bringing the micro-stamper down to transfer the desired flake within the targeted area.

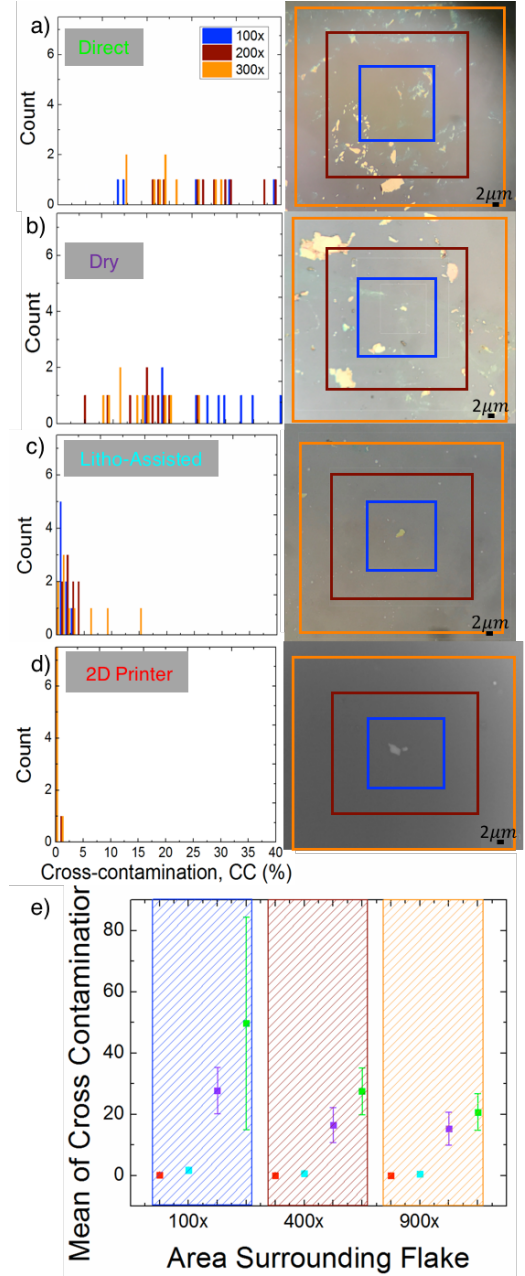

**Figure 2.** Quantitative data and analysis of the cross contamination (*CC*, see Eqn. (1) main text) on the substrate for each transfer method. In order to normalize our approach for varying 2D flake sizes, for each of the four transfer options investigated our methodology is as follow; (i) the size of the flake that is transferred into the target area ($A_{flake}$) is measured, (ii) the area box ($A_{box}$) in which cross contamination is measured is taken as factors ($f$ = 100x, 400x, and 900x) times that of the transferred flake area (i.e. $A_{box} = f \times A_{flake}$), and (iii) 10 flakes are transferred for each of the four transfer methods and a histogram is created along with a representative optical microscope image (a-d). (a) Direct. (b) Dry. (c) Litho-Dry. (d) 2D Printer. (e) Fitting the *CC* histograms with a normal distribution we obtain the mean (data point) and standard deviation (error bar) for all four methods (color-coded). The shaded region corresponds to the three selected factors, *f*. Our results show that for small areas near the transferred flake (e.g. $f$ = 100x) the cross contamination is about 30% and more than 50% for the Dry and Direct methods, respectively. In contrast *CC* is vanishing for the 2D Printer method and independent of $A_{box}$. The decline in CC (including its error) for the Direct and Dry methods is understood as a result of spatial-clustering introduced when obtaining a 2D material PDMS source (e.g. exfoliation was used).

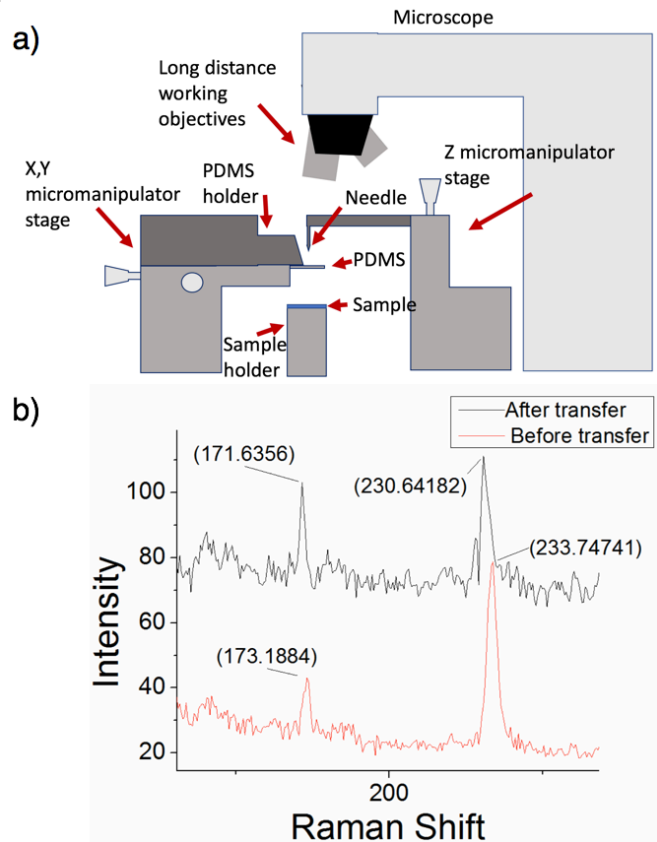

**Figure 3.** Schematic of the 2D printer is illustrated and described along with a Raman spectroscopy graph of the flake before and after the transfer process. (a) Schematic diagram of the 2D printer. The setup of the 2D printer consists of a micromanipulator stage that rotates. The PDMS gel is held above the substrate with the PDMS clipped together on both sides. The gel is stretched out until it is taut to increase the likelihood of transferring a single flake. A micro-stamper (e.g. needle) is placed over the gel with an adjustable three-axis micromanipulator. (b) Raman spectroscopy graph of before and after 2D Printer transfer of a TMD flake showing the preservation of the material quality upon micro stamping using the 2D Printer method. 2D Material is $MoTe_2$.

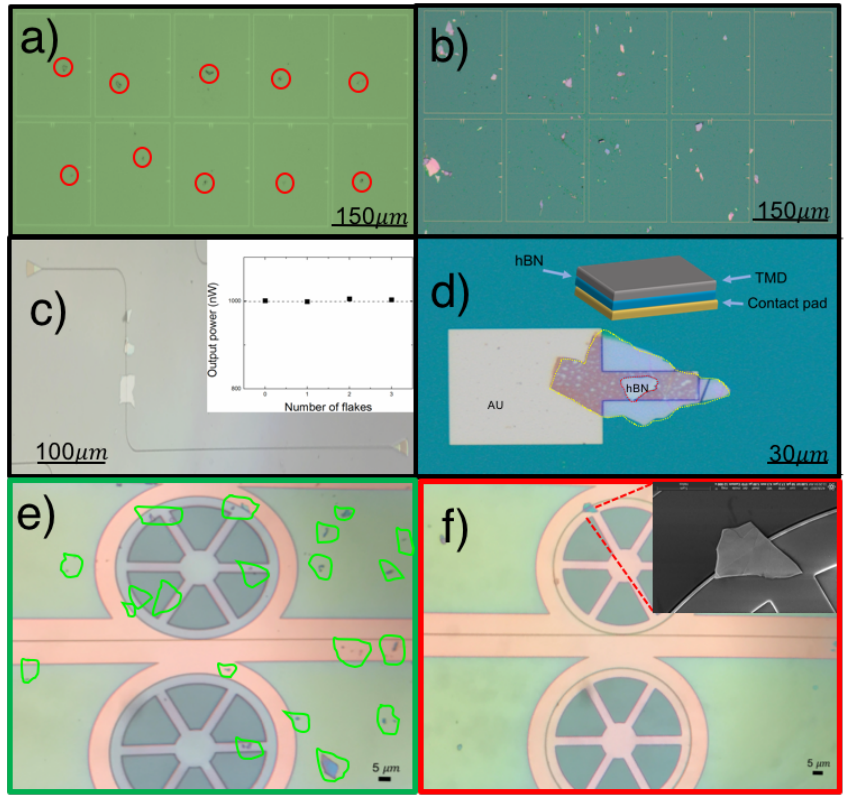

**Figure 4.** Applications based demonstration of 2D Printer transfer method at work. (a) 5x2 array of flakes on slot waveguides using 2D Printer method. The flakes are placed inside the boxes with no observable zero cross contamination. Towards realizing electronic or photonic devices, heterostructures can be built in this fashion as demonstrated in panel d), or waveguide can be placed onto the transferred flake. (b) 5x2 array of flakes on slot waveguides using Dry method. The high amount of cross contamination near targeted transferred flakes renders this method impracticable if more than a single (or few) devices are being fabricated. (c) Demonstration of hybrid integration of three TMDs being transferred onto a silicon photonic waveguide using the 2D Printer method showing an unchanged power output of the waveguide. This indicates that placing the 2D materials onto waveguides is a benign method. (d) Fabricated heterostructure demonstrating the ability to precisely transfer flakes of different materials towards building functional devices. Here the TMD (MoTe$_2$) could be electrostatically gated if contacted against the bottom gold pad. (e) The Direct transfer method is used to transfer a flake onto a ring resonator. However the lack of selectivity of this method transfers more than the target flake, depending on the source quality. If exfoliation is used as a source, the to-be-stamped PDMS usually contains clusters of 2D materials leading to a high amount of cross contamination potentially ruining neighboring devices or waveguides. On sensitive and costly chips (e.g. such as on tape-outs as done here using silicon photonics), this method would not be viable considering the amount of cross contamination that is unpreventable as well as the randomized approach to get a flake onto the target area. (f) The 2D Printer method is used to precisely transfer a single flake onto a ring resonator. The optical microscope image shows that the flake is successfully transferred onto a single micro ring resonator with no cross contamination. Inset: SEM image of the transferred flake on the microring ring resonator.